\documentclass[prb,twocolumn,showpacs,preprintnumbers,amsmath,amssymb]{revtex4}
\usepackage{graphicx}
\usepackage{dcolumn}
\usepackage{bm}
\usepackage{wasysym}

\newcommand{\s}{{\sigma}}

\newcommand{\be}{\begin{eqnarray}}
\newcommand{\ee}{\end{eqnarray}}
\newcommand{\nn}{\nonumber\\}
\newcommand{\Eq}[1]{Eq.~(\ref{#1})}

\begin{document}
\title{Geometric effects on $T$-breaking in $p+ip$ and $d+id$ superconductors}\author{J.~E.~Moore}

\affiliation{Department of Physics, University of California,
Berkeley, CA 94720} \affiliation{Materials Sciences Division,
Lawrence Berkeley National Laboratory, Berkeley, CA 94720}

\author{D.-H.~Lee}
\affiliation{Department of Physics, University of California,
Berkeley, CA 94720} \affiliation{Materials Sciences Division,
Lawrence Berkeley National Laboratory, Berkeley, CA 94720}

\date{\today}
\begin{abstract}
Superconducting order parameters that change phase around the
Fermi surface modify Josephson tunneling behavior, as in the
phase-sensitive measurements that confirmed $d$ order in the
cuprates.  This paper studies Josephson coupling when the
individual grains break time-reversal symmetry; the specific cases
considered are $p \pm ip$ and $d\pm id$, which may appear in
Sr$_2$RuO$_4$ and Na$_x$CoO$_2 \cdot $(H$_2$O)$_y$ respectively.
$T$-breaking order parameters lead to frustrating phases when not all
grains have the same sign of time-reversal symmetry breaking, and
the effects of these frustrating phases depend sensitively on
geometry for 2D arrays of coupled grains. These systems can show
perfect superconducting order with or without macroscopic
$T$-breaking.  The honeycomb lattice of superconducting grains has
a superconducting phase with no spontaneous breaking of $T$ but
instead power-law correlations.  The superconducting transition in
this case is driven by binding of fractional vortices, and the
zero-temperature criticality realizes a generalization of Baxter's
three-color model.
\end{abstract}
\maketitle
\section{Introduction}

Several materials are now believed to support superconducting
states where the order parameter $\Delta({\bf k})$ is
complex-valued and hence breaks time-reversal symmetry.  This
class includes the recently reported superconductor\cite{takada}
Na$_x$CoO$_2 \cdot $(H$_2$O)$_y$, which is a doped Mott insulator
since CoO$_2$ is a layered triangular-lattice spin-half
antiferromagnet. RVB-type mean-field theories\cite{baskaran}$^,$\cite{wang,kumar}
predict that the pairing symmetry should be $d\pm id$, and
successful observation of this state would lend strong support to
the RVB picture of unconventional superconductivity. $T$-breaking
superconducting pairing is also believed to occur in the layered material
Sr$_2$RuO$_4$\cite{mackenziemaeno}, where triplet Cooper pairs
have zero spin along the perpendicular direction ($S_z=0$) and two degenerate orbital states $p \pm ip$ with angular momentum $\pm \hbar$ in the perpendicular direction.

A major experimental success in the study of the cuprates was the
use of phase-sensitive measurement to confirm the $d_{x^2-y^2}$
pairing symmetry\cite{tsuei,vanharlingen}. In these experiments
the coupling between the superconducting order parameter and the
underlying crystalline lattice is used to design a frustrated loop
of weak links. (Each frustrated loop is associated with a ``flux''
which is the phase of the product of the Josephson coupling
constant around the loop.) Similar experiments may constitute a
powerful test of $p+ip$ and $d+id$ symmetries (section~\ref{secttwo}).

The main purpose of this paper is to demonstrate the rich physics
of arrays of Josephson junctions made up of superconductors with
$T$-breaking order parameters.  Throughout this paper we shall
focus on two dimensional arrays. In particular we shall show that
even in the absence of external magnetic field there is a
effective flux associated with each elementary loop that frustrate
the Josephson energy. This flux is determined by the Ising-like
time-reversal order parameters in the grains along that loop.  The
Ising order parameter specifies which of two degenerate states
related by time reversal exists in a particular grain: for
example, $d+id$ ($\sigma=+1$) or $d-id$ ($\sigma=-1$). The
classical Hamiltonian of such a system takes the Higgs-gauge
form
\be H_{HG} =
-E_J\sum_{<ij>}\cos(\phi_i-\phi_j-A_{ij}(\sigma_i,\sigma_j)).\label{hg}\ee
In the above $\phi_i$ is the phase of the superconducting order
parameter in the $i$th grain and $\sigma_i$ is the Ising-like time
reversal order parameter in that grain. The gauge field $A_{ij}$
is given by \be A_{ij}=k
\s_i\theta_i-k\s_j\theta_j,\label{gauge}\ee where $k=1$ for $p\pm
ip$ and $k=2$ for $d \pm id$, and $\theta_i$ is the angle between
the crystalline $y$-axis of the $i$th grain and the normal of the
interface from  $i$ to $j$. The existence of these discrete
(Ising) variables as well as the continuous superconducting phase
variables, and the fact the former affects the latter as a gauge
field, is the main difference between the problem we study here
and the usual frustrated/unfrustrated Josephson junction arrays.
Frustrated $s$ Josephson arrays have recently attracted attention
as a model for quantum computation\cite{ioffefeigelman,doucot}.
Direct coupling between local impurity spin variables and a
superconducting order parameter is discussed in\cite{henelius}.

Unlike in the real $d$ or $s$ case, Ginzburg-Landau theory for $T$-breaking
superconductors\cite{sigristueda} does not correctly give even the
phase diagram of Josephson arrays.  For example, the coupling
between superconducting phase and array geometry can induce
classical discrete $\mathbb{Z}_2$ or $\mathbb{Z}_3$ topological
order, where a fractional vortex of the superconducting phase is
bound to a defect of the local Ising variables that carries
fractional Ising quantum number. This is analogous to the Quantum
$\mathbb{Z}_2$ topological order that has attracted considerable
attention recently because of its appearance in quantum dimer and
other models\cite{senthilz2, moessnersondhi}.

Many of the nontrivial properties of Josephson arrays of $T$-breaking superconductors are
connected to models of frustrated magnets. For example, the
Ising configurations $\{\sigma_i\}$ that leave the interaction
among the phase variables unfrustrated are equivalent to the
ground-state spin configurations of certain frustrated magnets. On the
honeycomb lattice, the constraint of no
frustration for $p\pm ip$ or $d\pm id$ superconductors becomes the same as that in the``three-coloring problem'' \cite{baxter}, studied earlier because of its relation to
classical frustrated antiferromagnets on the kagom\'e
lattice\cite{chalkershender,readunpub,henleyunpub,huse,chandra}.
As in the XY antiferromagnet on the kagom\'e, the zero-temperature
superconducting phase has long-range order of $\psi^3$ but not
$\psi$, where $\psi$ is the Cooper pair operator.  That is, the phase of $\Delta({\bf k}_0)$ for some fixed ${\bf k}_0$ is not uniform throughout the array in the low-temperature phase, but varies only by integer multiples of $2 \pi / 3$.

Another
example of our results is that, on the square lattice, the Ising configurations that do not
cause phase frustration for $p+ip$ superconductors are the same as the ground
state gauge field configurations of a $\mathbb{Z}_2$ gauge theory.  The low-temperature state is like a nematic in that $\psi^2$ but not $\psi$ has an expectation value.

This paper is organized as follows. In section II we calculate the
Josephson tunneling amplitude between two $T$-breaking superconductors and
from microscopic considerations. In sections III we discuss some
general features of the Higgs-gauge model given in \Eq{hg}.  In
sections IV through VI we study the novel physics of \Eq{hg} on
different lattices. This corresponds to Josephson arrays of $T$-breaking
superconductors in different geometries. Section VII discusses the
entropic contribution to the interaction free energy between
fractional vortices. Finally in section VIII we summarize the
main conclusions and some connections to other recent work.

\section{The Josephson coupling between two complex superconductors}

\label{secttwo}

Consider the tunneling Hamiltonian between two superconducting domains 1 and
2 \be H_t = - \sum_{k_\parallel,k_\perp,k^\prime_\perp}
\left(t(k_\parallel,k_\perp,k^\prime_\perp)
c^{\dagger}_{1;k_\parallel,k_\perp}
c_{2;k_\parallel,k^\prime_\perp} + h.c.\right) \ee Here the
momentum parallel to the interface ($k_\parallel$) is conserved by
the tunneling process, while the perpendicular momentum
($k_\perp$) is not, as appropriate for a sharp boundary.

Let us assume the crystalline $y$-axis of the two domains are
oriented at angles $\theta_1$ and $\theta_2$ with respect to the
interface normal. The Josephson coupling between these two domains
can be calculated from second-order perturbation theory. The
result is \be E_{12}(\theta_1,\theta_2;\phi_1,\phi_2) &&=
-\sum_{k_\parallel,k_\perp,k'_\perp}{|t(k_\parallel,k_\perp,k'_\perp)|^2
\over 2\Delta}\nn\times && \Big[
\Delta_1(k_\parallel,k_\perp)\Delta^{*}_2(k_\parallel,k^\prime_\perp)+
c.c.\Big], \label{ej}\ee where ${\Delta}$ is the superconducting
order parameter. In deriving the above we have used the fact that
$t(-k_\parallel,-k_\perp,-k'_\perp)=t^*(k_\parallel,k_\perp,k'_\perp).$

Now we specialize to the $d\pm id$ case.  With respect to the individual crystalline axis $(x,y)$ the $d+id$ order parameter has the form
\be\Delta(k_x,k_y) =
\Delta_0e^{i\phi}(k_x + i\s k_y)^2, \label{odp}\ee
where $\phi$ is
the phase of the superconductor order parameter and $\s$ is the
time reversal order parameter\cite{note}.  When transformed to the
interface coordinate system we have \be
&&\Delta_1(k_\parallel,k_\perp)= \Delta_0e^{2
i\s_1\theta_1}e^{i\phi_1}(k_\parallel+i\s_1k_\perp)^2\nn
&&\Delta_2(k_\parallel,k_\perp)=\Delta_0 e^{2 i\s_2
\theta_2}e^{i\phi_2}(k_\parallel+i\s_2k_\perp)^2.\label{rotated}\ee
Substituting \Eq{rotated} into \Eq{ej} we obtain \be &&E_{12}=-E_J
\cos(\phi_1-\phi_2+A_{12}(\s_1,\s_2))\nn&&A_{12}(\s_1,\s_2)=2\s_1\theta_1-2\s_2\theta_2.\label{single}\ee

In the above \be
E_J&&=\Delta_0^2\sum_{k_\parallel,k_\perp,k'_\perp}{|t(k_\parallel,k_\perp,k'_\perp)|^2
\over  2\Delta}(k_\parallel+i\s_1k_\perp)^2\nn&&\times(k_\parallel-i\s_2k'_\perp)^2.\ee
Note that when $|t(k_\parallel,k_\perp,k'_\perp)|^2$ is even upon
reversing $k_\perp$ and $k'_\perp$ (as should be the case for
local tunnelling) $E_J$ is independent of $\s_1$ and $\s_2$.

Similar expressions can be obtained for $p+ip$ pairing except in this case
$A_{12}(\s_1,\s_2)=\s_1\theta_1-\s_2\theta_2.$  For a straight boundary, $\theta_2 = \theta_1 + \pi$.  This $\pi$ phase between opposite orientations is unimportant for even angular momentum order parameters (e.g., $d+id$) but significant for odd angular momentum order parameters (e.g., $p+ip$).


\section{Josephson junction arrays of $T$-breaking superconductors}

\label{secthree}

We now show from \Eq{single} that the classical Hamiltonian for an
array of $p+ip$ or $d+id$ Josephson junctions is given by the
gauge theory \Eq{hg} and \Eq{gauge}.  The Josephson Hamiltonian in
terms of gauge-invariant phase differences across each junction is
\be H = - E_J \sum_{\langle ij \rangle} \cos \phi_{ij}. \ee These
phase differences are related to the superconducting phases on the
grains by $\phi_{ij} = \phi_i - \phi_j - A_{ij}$ with $A_{ij}=k
\s_i\theta_i-k\s_j\theta_j$. The partition function of the system
is \be Z=\prod_j\sum_{\s_j}\int d\phi_j e^{\beta E_J
\sum_{ij}\cos(\phi_i-\phi_j-A_{ij}(\s_i,\s_j))}.\ee This theory
has a classical compact $U(1)$ gauge invariance under \be \phi_i
\rightarrow \phi_i + \chi_i, \quad A_{ij} \rightarrow A_{ij} +
\chi_i - \chi_j. \ee Standard duality
transformation\cite{kosterlitzthouless,jose,dhl} applied to the
above model gives the vortex partition function \be
Z\propto\sum_{\{\s_j\}}\sum_{\{n_a\}}e^{-\Delta_c\sum_a
(n_a-\Phi_a)^2
-\sum_{a,b}(n_a-\Phi_a)G_{ab}(n_b-\Phi_b)}.\label{dual}\ee In the
above $a, b$ label sites of the dual lattice, the $n_a$ are
integers, and the flux is \be\Phi_a={1\over 2\pi}\sum_{<ij>\in
a}A_{ij}(\s_i,\s_j).\label{flux}\ee The fugacity of vortices is
$e^{-\Delta_c n_a^2}$ and $G_{ab}$ is a logarithmic interaction
between the vortices.  In this form, the problem becomes a 2D
vortex gas with annealed flux variables. Each flux configuration
is weighted by a purely entropic term that measures how many Ising
configurations correspond to it. This type of problem has been
encountered previously in Ref.\cite{dhl}.

\section{Triangular lattice: confinement}

\label{tricrit}

\begin{figure}
\includegraphics[width=2.5in]{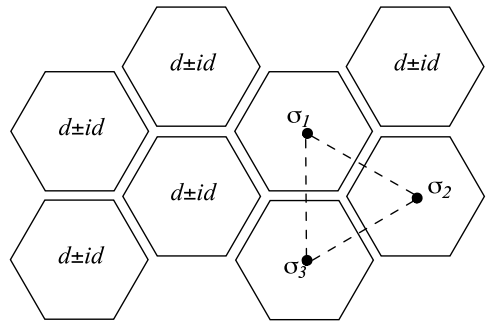}
\caption{A triangular lattice of $d \pm id$ superconducting grains connected by Josephson tunneling.  The state of grain $j$ is determined by an Ising variable
$\s_j = \pm 1$ and a superconducting phase.
}
\label{figtwo}
\end{figure}

As a first step in understanding how the gauge phases in \Eq{hg}
affect macroscopic ordering, we consider a regular
triangular lattice of $d+id$ or $p+ip$ grains (Fig.~\ref{figtwo}).
For an elementary plaquette with three corners $1,2,3$, it is
simple to show that the sum of $A_{ij}$ gives
rise to flux
\be \Phi_a = {k \over 3} (\s_1 + \s_2 + \s_3),~~\mod 1.
\label{topcons} \ee
The constant $k$ is 1 for $p+ip$ and 2 for
$d+id$.  The ground state configuration of $\{\s_i\}$ satisfies
the constraint that $\Phi_a$ be an integer for all plaquettes ($a$ runs
through a honeycomb lattice). The only way to satisfy this
constraint is for all $\s_i$ to take the same value $+1$ or $-1$.

\begin{figure}
\includegraphics[width=2.0in]{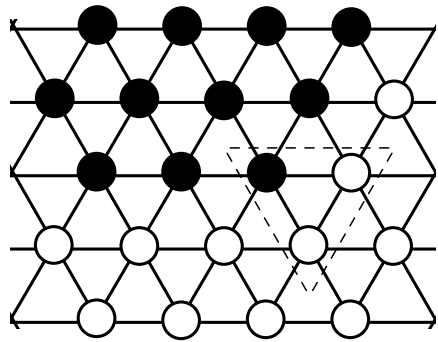}
\caption{Domain wall kink on the triangular lattice: there is a localized fractional flux $\Phi_0/3$ from the plaquette inside the dotted triangle.  Here solid (open) circles correspond to spin up (down).
}
\label{figflux}
\end{figure}

Flipping a single $\s$ creates six neighboring plaquettes with
fractional $\Phi_a=\pm 2k/3$, so the total flux is integral. It is
possible to create local fractional fluxes at domain wall
``kinks'': these are corners in a line separating a region of
Ising spin +1 from one of Ising spin -1 (Fig.~\ref{figflux}).
These fractional fluxes are connected by a defect line which costs
finite energy per unit length.  The entropic term is quite strict
for the triangular lattice, since any flux configuration corresponds
to at most two spin configurations.  As a result, the fractional
vortices are confined. To summarize, the triangular case has
perfect order of both the phase and Ising variables at $T=0$.We
expect that the finite-temperature superconducting transition will
be of the conventional KT type, triggered by the unbinding of
integer vortices.

It turns out that the triangular lattice of coupled grains is
special in its high degree of order at low temperature.  The
square lattice for $d\pm id$ grains is at the opposite extreme, as
shown in the following section: there is no coupling between the
Ising variables and the superconducting phase.  The zero-temperature degeneracy for the square lattice is thus given by a global $U(1)$ for
the superconducting phase times a factor of 2 per site from the
Ising variables.

The case of $p\pm ip$ superconductors on the square lattice is
quite complex and is shown below to be connected to the existence
of a $\mathbb{Z}_2$ topological order.  We find that $p \pm ip$ superconductors on the square lattice support localized $\pi$-vortices with half the enclosed flux of an ordinary
superconducting vortex, and these vortices drive the superconducting transition.

\section{Square lattice: deconfinement}
Square arrays of both $p+ip$ and $d+id$ superconductors show
dramatically different behavior from the triangular lattice
discussed above: both have no order of the Ising $T$ variables
even at zero temperature when only the Josephson coupling is
considered, i.e., only single-Cooper-pair processes are included
in the model.

The gauge flux $\Phi_a$ that pierces an elementary
plaquette in the square lattice is given by
\be \Phi_a={k \over 4}
(\s_1 + \s_2 + \s_3 + \s_4)~~\mod 1. \label{z2cons}\ee
Again $k=1$ for $p+ip$ and $k=2$ for $d+id$. For $d+id$ the flux generated by
the Ising variables is ${1\over 2}\sum_{\square} \s_i$, which is
always an integer. Since integer gauge flux can always be gauged
away, the problem decouples into a square-lattice XY model and a
non-interacting Ising model.

\begin{figure}
\includegraphics[width=1.5in]{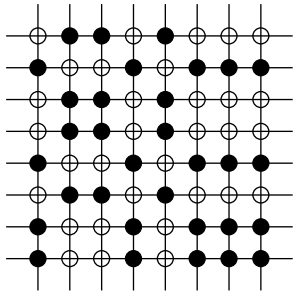}
\caption{A typical ground state configuration of Ising variables
(open or filled circles) for $p\pm ip$ superconductors.  Note that
fixing one row and one column determines the rest of the lattice
via the constraint of 0, 2, 4 up-spins per square. } \label{figsqr}
\end{figure}

The case of $p+ip$ arrays is more interesting. In this case
$\Phi_a={1\over 4}\sum_{\square} \s_i$ and hence can be fractional.
Indeed, for 0, 2, or 4 up spins per square $\Phi_a$ is an integer,
while for 1 or 3 up spins $\Phi_a$ is $\frac{1}{2}$ plus an integer. In the
ground state, the Ising spin configurations must satisfy the
constraint that there are even number of up spins in every
plaquette. In a system with open boundary conditions and size $M$
by $N$, the number of such Ising configurations is $2^M\times
2^N$. The first factor corresponds to the number of ways to fix all
the spins on one horizontal line, and the second factor the number
of ways to fix all the spins on one vertical line. Once those
spins are fixed, the remaining spins are uniquely determined by
the constraint (Fig.~\ref{figsqr}).  This implies that, although
the above system appears two-dimensional, its ground-state entropy
grows as the system's linear size rather than its area.

The ground-state constraint of 0, 2, or 4 up-spins per square
corresponds to a degenerate case of the eight-vertex
model\cite{baxterbook}, 
which can be mapped onto the six-vertex model\cite{lieb6} using a
duality transformation\cite{fanwu}.  The point described by
${1\over 4}\sum_{\square} \s_i$ integral is quite special: it
divides a region of the phase diagram with extensive
(two-dimensional) entropy from a region with finite
(zero-dimensional) entropy.

The ground states for $p\pm ip$ superconductors on the square lattice are the
same as those of the Ising lattice gauge model
\be H = -
K \sum_{i,j} \sigma_{(i,j)} \sigma_{(i+1,j)} \sigma_{(i+1,j+1)}
\sigma_{(i+1,j)}, \label{z2gauge} \ee
since the product around a
plaquette takes the value $+1$ for $0, 2, 4$ up spins, and $-1$
otherwise. A plaquette with $1$ or $3$ up spins generates a gauge
flux. Whether such defective plaquettes are entropically confined
is equivalent to whether the the $\pi$-gauge flux is confined in
the gauge theory when $K\rightarrow 0$.  Our previous degeneracy counting implies that the $\pi$ gauge flux would completely deconfined if the entropic term were the only
contribution to its free energy. In reality, in addition to the
entropic contribution, there is a phase stiffness contribution to
the interaction between two fractional gauge flux. Indeed, when
$\Phi_a$ is a fraction, it cannot be screened by integer vortices $n_a$.
As the result, two fractional fluxes $\Phi_a$ separated by a large distance $R$
cost free energy $\sim (\pi K / 4) \log R$, where $K$ is the renormalized phase
stiffness. This logarithmic interaction binds half-integer
vortices at low temperatures.

The half-integer vortices are relevant at the ordinary (integer
vortex) XY transition because their scaling dimension is four
times smaller than that of the integer vortices, which are
marginal at the transition.  It follows that as the temperature is
raised above a critical value, the half-integer vortices unbind
and the superconductivity is lost.  The superconducting phase
below this temperature has ``nematic'' order in that there is
long-range order in the square $\psi^2$ of the Cooper pair
operator, rather than in $\psi$.

 A defective plaquette can be viewed as a $\pi$-flux bound to a fractional Ising charge.  The Ising charge
associated with a given plaquette is defined as \be
Q_I={(N_{\uparrow,\square}-2) \over 2}, \ee where
$N_{\uparrow,\square}$ is the number of up spins in the plaquette.
Any plaquette satisfying the 0, 2, 4 constraint has an integral
$Q_I$. An isolated plaquette with 1 or 3 up spins has half-integer
$Q_I$. Since the Josephson energy is only sensitive to whether the
$Q_I$ is integer or half-integer, one should regard $Q_I$ as
defined modulo integer since two half-integer $Q_I$ cancel. Local
constraints like (\ref{z2cons}) were used
in\cite{motrunichsenthil} to generate a gauge theory, but required
a somewhat artificial charging energy that is associated with
plaquettes, not with single grains or with long-range
interactions.  We have shown that geometric phases in Josephson
junction arrays naturally generate such a plaquette constraint,
and in turn a classical two-dimensional $\mathbb{Z}_2$ gauge
theory.

\section{Honeycomb lattice: criticality}
\label{hexsect} We now study coupled $p \pm ip$ and $d \pm id$
grains on the honeycomb lattice and show that the essential
low-temperature physics is considerably modified from the
triangular and square cases studied in the preceding sections.
This model will be shown to have an extensive entropy down to zero
temperature with no macroscopic order of the Ising sign variable.
Hence even though locally (i.e. in each superconducting grain)
T-breaking superconductivity is favored, globally there is no
time-reversal symmetry breaking; there is instead a critical state at zero temperature.

For the honeycomb lattice, the gauge flux in each hexagon is given
by \be \Phi_a={k \over 6} (\s_1 + \s_2 + \s_3+\s_4+\s_5+\s_6),
~~\mod 1. \ee Hence a plaquette with zero, three, or six up-spins
has integer $\Phi_a$, and does not cause any frustration in the
Josephson energy. Thus the lowest energy Ising spin configurations
satisfy the 0, 3, 6 constraint in every plaquette. The number of
such configurations can be found exactly in the thermodynamic
limit by a mapping\cite{difrancesco1} (section \ref{critsect})
onto the three-color model solved exactly by Baxter\cite{baxter}.
The result gives an entropy per hexagon equal to $0.379114 k_B$.
The critical properties of this three-color model are briefly
reviewed in section VII.

Ising configurations that violate the 0, 3, 6 constraint lead to fractional gauge fluxes.
The number of Ising configurations consistent with a distribution
of fractional gauge fluxes is different from that of the ground
state degeneracy discussed above. The logarithm of the ratio
between these degeneracies gives rise to a entropic contribution
to the interaction free energy between fractional gauge fluxes. To
compute such interaction we need to consider a correlation
function that does not have an analogue in the three-color model: this is the main subject of section VII.  Here we explore the physical consequences of the entropic interaction.

To be more precise let us return to \Eq{dual}. For a fixed
distribution $\{\Phi_a\}$ ($-1/2\le\Phi_a< 1/2$) the entropic
contribution to the dimensionless free energy is given by
\be e^{-\Delta
F(\{\Phi_a\})}={\sum_{\{\s_j\}} \prod_a {\tilde \delta}\Big(\Phi_a-{k
\over 6} \sum_{j\in a}\s_j \Big)\over\sum_{\{\s_j\}} \prod_a
{\tilde \delta}\Big({k \over 6} \sum_{j\in
a}\s_j\Big)}.\label{ratio}\ee
Here the function ${\tilde \delta}(x)$ is 1 if $x \in \mathbb{Z}$, 0 otherwise.  After the summation over the
Ising variables, \Eq{dual} becomes
\be
Z&&\propto\sum_{\{n_a\}}e^{-S(\{n_a,\Phi_a\})},\ee where \be S&&=
\Delta_c\sum_a (n_a-\Phi_a)^2
+\sum_{a,b}(n_a-\Phi_a)G_{ab}(n_b-\Phi_b)\nn&&+\Delta
F(\{\Phi_a\}).\label{dual1}\ee The question is what effect, if
any, does $\Delta F(\{\Phi_a\})$ have on the thermodynamics of the
vortices?

Section~\ref{critsect} shows that the probability $e^{-\Delta F}$ associated
with two $\Phi_a=\pm 1/3$  defect plaquettes of opposite
fractional flux falls off with distance approximately as
$r^{-4/9}$. This implies a free energy cost of
\be \Delta F = 4~
\Phi_a \log|r_a-r_b| \Phi_b. \label{entro}\ee

Imagine the stiffness $K$ is tuned so that the system is exactly
at the critical point where integer vortices proliferate if
fractional vortices are strictly forbidden.  Then the vortex
fugacity operator is marginal (has scaling dimension 2). From the
Gaussian spin-wave theory vortices of fractional charge 1/3 have
scaling dimension $\Delta = 2/9$. However, this analysis ignores
the entropic contribution to the free energy. Including that
effect from \Eq{entro} implies the total scaling dimension of fractional 1/3
vortices at the (integer vortex unbinding) KT point
\be \Delta =
2/9+2/9=4/9.  \ee
This indicates that fractional
vortices are in fact relevant, and most likely it is their
unbinding that drives the superconductor-normal XY-type
transition.  The superconducting phase is then a ``triatic'' with long-range order in the cube of the Cooper pair operator $\psi^3$ rather than $\psi$.

There are several corrections in realistic systems (quenched
disorder, multiple-Cooper-pair processes, etc.) that will split
the extensive zero-temperature entropy, as discussed briefly in
the final section and in a forthcoming paper by Castelnovo, Pujol,
and Chamon\cite{chamonunpub}.  However, these effects can be
expected to onset at temperatures much less than the Josephson
coupling ($kT \ll E_J$), so experiments on the honeycomb lattice
at the lowest accessible temperatures are likely to see an
extensive entropy and critical correlations of the time-reversal
order parameter.  The following section obtains these critical
properties using the connection to the three-color model.

\section{The $T=0$ critical theory on the honeycomb lattice}
\label{critsect}As  shown in section~\ref{hexsect}, the ground state
configurations of the Ising spins have 0, 3, or 6 up-spins on each
hexagon. The entropy per site of this ``0, 3, 6-model'' was
earlier shown\cite{difrancesco1,difrancesco2} to be equal to that
of the ``three-color problem'' solved by Baxter. In the
three-color problem, one assigns one of three colors (say red,
green, blue) to each bond of the honeycomb lattice, with the
requirement that all three colors meet at each site. A quantity
called chirality can be defined in the following way. At each site
there are three bonds each being colored differently. Start with
the red bond and turn clockwise: if the sequence of colors
encountered is RGB we assign chirality $\s=+1$ to that site. If
the sequence is RBG we assign $\s=-1$. In this way each coloring
pattern is uniquely associated with an Ising configuration. The 0,
3, 6 constraint ensures that one returns to the original color
after going around a plaquette (hexagon). Since the colors are
uniquely determined from the Ising variables once one bond is
fixed, the entropy of the two models differs by a global factor of
3. Baxter has shown that the entropy per hexagon is \cite{baxter}
\be S = \log\left({2^2 \over 1 \cdot 3} {5^2 \over 4 \cdot 6} {8^2
\over 7 \cdot 9} \cdots \right) \approx 0.37905. \ee

The three-color model is an example of zero temperature models
that have critical correlations\cite{berker}, and many of its
critical properties are known exactly. To determine these
properties it is useful to represent the three-color model as a
two-component height model (on the dual, triangular, lattice) as
follows\cite{readunpub,huse,kondev}. Consider a given site on one
sublattice (say $A$), and the three colored bonds stemming from
it.  If we cross a given bond
 in a clockwise fashion with respect to the site, the following increment to the high
variable is assigned: \be \Delta {\bf h}
&&=(1,0)~~red\nn&&=(-1/2,\sqrt{3}/2)~~blue\nn&&=
(-1/2,-\sqrt{3}/2)~~green \label{heightvals} \ee This assignment
is self-consistent since the three-color constraint ensures the
total height increment is 0 after one completes a loop.  The height increment gets a factor of $-1$ if the crossing is counterclockwise with respect to sublattice $A$.  The
long-wavelength effective theory for this problem is given by
\be
F = \int \left(\frac{K}{2} (\nabla {\bf h})^2 + V({\bf
h})\right)\,d^2x, \label{gaussian} \ee
where the potential $V$ has the periodicity of a honeycomb lattice of
side $1/\sqrt{3}$.  The stiffness $K$ is determined by the requirement that the locking potential $V$ be marginal.

In this theory the allowed vertex operators have the form ${\hat
O}_{\bf \alpha}(x) = \exp(i \alpha_i h_i(x))$ where ${\bf \alpha}$
lies on the reciprocal lattice of $V$.  The scaling dimension
$\Delta$ of such an operator, which determines its correlations
through $\langle {\hat O}_{\bf \alpha}(0) {\bar {\hat O}}_{\bf
\alpha}({\bf x})\rangle \sim r^{-2 \Delta}$, is given by $\Delta =
\alpha^2/(3\pi)$. Knowing the critical theory of the three color
problem it is possible to compute the entropy associated with
certain localized defects in the color assignment. These defects
appear as vortex operators; among them the most relevant one has
scaling dimension $\Delta=1/2$. This information allows us to
determine the full critical three-color theory.  This critical
theory (\ref{gaussian}) has a hidden $SU(3)$
symmetry\cite{readunpub}. The field-theory representation of the
$SU(3)$ symmetry of (\ref{gaussian}) is given in\cite{kondev}.

Unfortunately knowledge about the three-color model does
not enable us to compute the entropic contribution to the
interaction between fractional gauge fluxes as in \Eq{entro}. This
is because a single plaquette with the wrong number of up Ising
spins spoils the mapping to the three-color model. For example, on
circling a defect plaquette with only 2 Ising spins, the colors
wind up cyclically permuted. To find the entropic interaction
between a pair of fractional vortices with opposite vorticity, we
numerically computed the ratio in \Eq{ratio}.

We have performed numerical transfer-matrix calculations on strips
and cylinders of up to 9 hexagons width/circumference.  The
entropy for the ground state is calculated as follows:
configurations allowed by the constraint (0, 3, or 6 up-spins on
each hexagon) are assigned $\beta E = 0$, while forbidden
configurations have $\beta E = \infty$.  Let the length of the
strip be $N$: the entropy is given by \be S = \log Z = \log (T^N) =
N \log \lambda_1, \ee where $\lambda_1$ is the largest eigenvalue
(assumed nondegenerate) of the transfer matrix $T$. The entropy
per hexagon of the 2D system, if extensive, can then be estimated
as \be S/\hexagon = \lim_{W\rightarrow \infty} {\log \lambda_1(W)
\over W}, \ee where $W$ is the width of the strip. Table I shows
the resulting estimates with periodic and open boundary conditions
for strips of various widths.

\begin{table}
\begin{tabular*}{3 in}{@{\extracolsep{\fill}} c   c  c c c}
Width&\multicolumn{2}{c}{Estimated $S/\hexagon$}&$\Delta_d$&$\Delta_c$\\[0.5ex]
\hline
&Open&Periodic&\\
\hline\hline\\
1&1.05&\\
2&0.703&\\
3&0.590&$
0.462$&0.210\\
4&0.534& \\
5&0.502&\\
6&0.480&0.403&0.231&1.90\\
7&0.465&\\
8&0.454&\\
9&&0.390&0.235&1.91\\
$\vdots$&$\vdots$&$\vdots$\\
\hline
Theory ($\infty$) &0.379114&0.379114&&2\\
\end{tabular*}
\caption{Results of numerical transfer-matrix calculations: estimates of $T=0$ entropy and correlations.  The correlation between two defect plaquettes of opposite sign (e.g., one with 2 spins up and one with 4) falls off as $r^{-2\Delta_d}$, and the chirality correlation falls off as $r^{-2 \Delta_c}$.}
\end{table}

The transfer matrix can also be used to calculate the entropic
interaction between a pair of plaquettes with opposite flux on a
cylinder.  These defect plaquettes must be on the same sublattice
for a nonzero correlation.  We use the usual
procedure\cite{cardybook} for extracting scaling dimensions at a
critical point from calculations on the cylinder, based on the
hypothesis of conformal invariance.  The relative probability to
have positive and negative defect plaquettes at fixed locations,
described by transfer matrices $T_+$ and $T_-$, separated by $n$
defect-free rings around the cylinder, is \be p(n)={Z(n) \over Z}
= \lim_{N\rightarrow \infty} {{\rm Tr}\ T^N T_+ T^n T_- T^N \over
{\rm Tr}\ T^{2N+n+2}}. \ee If this function in the continuum limit
is a correlation function of a critical theory in 2D, then
conformal invariance predicts that correlations along the cylinder
at large distances $r \gg L$, $L$ the circumference of the
cylinder, should fall off as \be p(r) \sim e^{-r/\xi},\quad \xi =
{L \over 2 \pi \Delta}. \ee As seen in Table I, this
characteristic dependence is indeed observed for both the
fractional-vortex-defect correlator and the Ising correlator.

The
numerical results suggest that \be e^{-\Delta F(r_a,r_b)}\sim {1\over
|r_a-r_b|^{0.47}}.\label{result}\ee This implies that the
fractional vortices corresponds to an operator with scaling
dimension $\Delta \approx 0.235$ in the 0, 3, 6 model. Since the
lowest two scaling dimensions that appear in the three-color model
are $\Delta=1/2$ and $\Delta = 2/3$ which are very different from
$0.235$, it is apparent that the critical theory of the 0, 3, 6
model has a different operator content than the three-color
model.

Such a critical theory must have the same correlators of height
variables as the three-color model, and yet have a different set
of defect operators, including the defective plaquette with
$\Delta \approx 0.23$.  One possibility is the ``orbifold'' dual
of the above three-color model theory obtained by identifying
heights that differ by a $\mathbb{Z}_3$ transformation $h_1+ i h_2
\rightarrow \exp(2 i \pi n / 3) (h_1 + i h_2)$.  The physical
significance of this identification is that the fractional
vortices, around which the height lattice is rotated by $2 \pi /
3$, are now allowed defects.  On the torus, there are now
different sectors of the theory corresponding to how the color
scheme is shifted upon passing from one boundary of the torus to
another.  We find that the orbifold ``twist'' operators that
connect these different sectors\cite{dulat} have scaling dimension
2/9 and may correspond to the fundamental plaquette defects in the
0, 3, 6-model.  However, this identification is quite indirect at
present and should be regarded as only one of several
possibilities for the critical theory.  This orbifolded $SU(3)$
theory with $c=2$ has previously been considered as a hypothetical
model of a quantum Hall edge\cite{skoulakis}, since the
non-orbifolded $SU(3)$ theory describes several real quantum Hall
edges\cite{kanefisherbigedge,mooreedge} in the chiral Luttinger
liquid approach\cite{wenrev1}.

Finally, \Eq{result} indicates that fractional vortices have
already become unbound when the integer vortex pairs disassociate.
To see that we imagine forbidding the fractional vortices and
tuning the stiffness  constant $K$ so that the integer vortex just
about to unbind. At that point the vortex fugacity operator is
marginal (has scaling dimension 2), and 1/3 vortices have scaling
dimension $\Delta_i = 2/9$. Adding the entropic interaction
(\Eq{result}) this yields the total scaling of the fractional
vortex fugacity operator \be \Delta = \Delta_i + \Delta_e =
2/9+2/9=4/9. \label{fluxeq} \ee This indicates that fractional
vortices are in fact relevant, and most likely it is their
unbinding that drives the superconductor-normal XY-type
transition.  The superconducting state has ``triatic'' order in
$\psi^3$ rather than in $\psi$, as in the ground state of the
antiferromagnetic XY model on the kagom\'e lattice\cite{huse}.

\section{Summary and other directions: dynamics and quenched disorder}

Our main goal in this paper was to demonstrate that due to the
coupling of the order parameter and the underlying crystalline
lattice, Josephson junction arrays made of time-reversal-symmetry-
breaking superconductors exhibit very much richer physics than
their $s$ or $d$ counterparts.  Phase-sensitive experiments such as
those in Ref.\cite{tsuei,vanharlingen} should be able to detect
the variable tunneling phases caused by a $T$-breaking order parameter. Regular
arrays of such grains give rise to realizations of Higgs-gauge
models, and are new examples of topological order with deep
connections to frustrated magnets. The ordering of the
time-reversal order parameter and the onset of superconductivity
depends intricately on the geometry of the arrays.

One can expect, on a qualitative level, that the existence of the
nontrivial phases studied above will also modify the behavior in
random polycrystalline samples. The behavior in the random case is
likely quite complex since random Josephson coupling, even without
the gauge phases we have studied, leads to the  ``dirty boson''
problem\cite{fisher,vishwanath}.  Allowing random local
orientation of the order parameter for $d$ superconductors has
been argued to induce a superparamagnetism
effect\cite{sigristrice} that will coexist with the spontaneous
orbital currents in the $d+id$ case. A more tractable model would
be to allow a quenched time-reversal-symmetry breaking field
coupling to the Ising variables, but keep uniform the strength of
the Josephson coupling.

The possibility of a direct coupling between the Ising variables
at higher order in the Josephson tunneling is studied in a
forthcoming paper\cite{chamonunpub}.  This model has phase
transitions from the critical phase with no additional interaction
to ordered ferromagnetic and antiferromagnetic phases.  Its
dynamics are of special interest, as the three-color model has
previously been studied as a model system for glassy
dynamics\cite{chakrabortykondev}.  Our realization of the
three-color model in arrays of complex superconducting grains may
prove to be experimentally less challenging than recent
efforts\cite{parkhuse} to build frustrated XY models on the
kagom\'e lattice using $s$ superconductors and a frustrating
magnetic field.
\section{acknowledgements}

J. E. M. acknowledges helpful discussions and correspondence with C. Chamon, P. Fendley, O. Ganor, J. Kondev, and S. Thomas, and the hospitality of the Aspen Center for Physics.  The authors were supported by NSF DMR-0238760 and the Hellman Foundation (J. E. M.), and DOE (D. H. L.).    This research used resources of the National Energy Research Scientific Computing Center, supported by DOE contract DE-AC03-76SF00098.

\bibliographystyle{apsrev}

\bibliography{../bigbib}

\end{document}